\documentclass[ reprint,nofootinbib,amsmath,amssymb,aps]{revtex4-1}
\usepackage{graphicx}
\usepackage{subfigure}
\usepackage{dcolumn}
\usepackage[colorlinks,linkcolor=magenta,anchorcolor=cyan,citecolor=blue]{hyperref}
\usepackage{bm}
\usepackage[multiple]{footmisc}

\def\be{\begin{equation}}
\def\ee{\end{equation}}
\def\ba{\begin{eqnarray}}
\def\ea{\end{eqnarray}}

 \def\w{\omega}  \def\y {\psi}           \def\k {\kappa}     \def\b {\beta}      
          \def\L {\Lambda}    \def\.{\cdot}

\begin{document}
\title{Restoring Strong Cosmic Censorship in Reissner-Nordström-de Sitter Black Holes via Non-Minimal Electromagnetic-Scalar Couplings}
\author{Jie Jiang${}^2$}
\email{jiejiang@mail.bnu.edu.cn}
\author{Jia Tan${}^1$}
\email{Corresponding author. jiatansust@163.com}
\affiliation{${}^1$Jiangsu Key Laboratory of Micro and Nano Heat Fluid Flow Technology and Energy Application, School
of Physical Science and Technology, Suzhou University of Science and Technology, Suzhou, 215009, China}
\affiliation{${}^2$College of Education for the Future, Beijing Normal University, Zhuhai, 519087, China}
\begin{abstract}
We investigate whether the Strong Cosmic Censorship (SCC) Conjecture can be reinstated in Reissner-Nordström-de Sitter (RNdS) black holes by introducing non-minimal couplings between the electromagnetic and scalar fields in Einstein-Maxwell-scalar (EMS) theory. By conducting numerical calculations, we find that the SCC can be restored within a specific range of the coupling constant. Notably, for a given value of the cosmological constant, there exists a critical coupling constant above which the EMS theory satisfies the SCC. These findings suggest that the non-minimal couplings between the electromagnetic and scalar fields may play a crucial role in the restoration of the SCC in RNdS spacetime.
\end{abstract}
\maketitle

\section{Introduction}

The Strong Cosmic Censorship Conjecture (SCC) represents a critical and fascinating issue in classical general relativity (GR). First proposed by Roger Penrose in 1969 \cite{Penrose:1969pc}, the conjecture asserts that the Cauchy horizon (CH) of a black hole must be singular and, consequently, inextendible into the future. Based on the principle of determinism in classical GR, the SCC contends that our universe's future is dictated by initial conditions specified on a Cauchy surface. A non-singular CH would result in an undetermined future and a non-deterministic spacetime, which is why the SCC is regarded as a foundational principle in classical GR.

Though there are multiple formulations of the SCC, the most contemporary version was introduced by Demetrios Christodoulou in 2008 \cite{Christodoulou:2008nj}. This formulation posits that the spacetime metric cannot extend beyond the CH in the form of weak solutions to the field equations. In essence, the CH must be a mass-inflation singularity, signifying that field perturbations must evolve to become divergent on the CH. As a highly complex issue, proving or disproving the SCC remains one of the most vital and demanding challenges in the realm of GR.

The validity of the Strong Cosmic Censorship Conjecture (SCC) has been confirmed in specific black hole solutions, including asymptotically flat Reissner-Nordström and Kerr black holes \cite{Simpson:1973ua,Poisson:1990eh,Dafermos:2003wr}. However, recent findings have identified violations of the SCC in the nearly extremal region of the Reissner-Nordström-de Sitter (RNdS) black hole \cite{Cardoso:2017soq}. These violations are attributed to the presence of cosmological horizons, which can suppress the blueshift effect responsible for the mass inflation singularity at the Cauchy horizon (CH) \cite{Brady:1996za,Costa:2014aia,Hintz:2015jkj,Costa:2016afl}. Following this, further violations have been discovered in the RNdS background by massless charged scalar fields, massless Dirac fields, and others \cite{Mo:2018nnu,Ge:2018vjq,Liu:2019rbq,Dias:2018ufh,Cardoso:2018nvb,Destounis:2018qnb,Liu:2019lon,Destounis:2020yav}.  Hollands et al. \cite{Hollands:2019whz} made progress in understanding the SCC in RNdS black holes, showing that non-singular initial quantum field data can restore the SCC. The review \cite{Kontou:2020bta} provides a comprehensive discussion on energy conditions in GR and quantum field theory. Moreover, investigations into the validity of the SCC for the RNdS black hole under classical perturbations of non-minimally coupled scalar fields in certain modified gravitational theories have been conducted \cite{Destounis:2019omd,Guo:2019tjy}. Within the framework of the Einstein-Maxwell-Scalar-Gauss-Bonnet (EMSG) theory, Ref. \cite{Sang:2022frw} has demonstrated that the SCC can be restored for RNdS black holes under specific parameter ranges when incorporating non-minimally coupled scalar field perturbations. This finding offers a fresh perspective for studying the SCC in modified gravitational theories.  The understanding of the SCC continues to be an active research area, bearing significant implications for the fundamental principles of classical GR and the ultimate fate of our universe.

The Einstein-Maxwell-Scalar (EMS) theory has attracted significant interest due to its non-minimal couplings between the electromagnetic and scalar fields and the resulting black hole solutions. These couplings have long been considered in the context of theories such as Kaluza-Klein and supergravity \cite{Gibbons:1987ps,Garfinkle:1990qj}, and more recently, in the context of black hole spontaneous scalarization, a strong gravity phase transition \cite{Damour:1993hw,Herdeiro:2018wub}. EMS models use a non-minimal coupling between the scalar field and Maxwell invariant to induce scalarization, requiring the presence of electric or magnetic charge. Studying the EMS models has led to the discovery that spontaneous scalarization of charged black holes occurs dynamically, resulting in scalarized, perturbatively stable black holes \cite{Herdeiro:2018wub,Fernandes:2019rez,Myung:2018jvi,Myung:2018vug,Myung:2019oua}.

Given the similarities between EMS and ESGB, it is worth exploring whether SCC can also be restored in EMS. Studying SCC in EMS is highly significant as it can provide valuable insights into the interplay between non-minimal couplings, spontaneous scalarization, and the fundamental principles of classical GR \cite{Brady:1996za,Cardoso:2017soq}. Understanding whether SCC can be restored in EMS may ultimately deepen our knowledge of black holes and the underlying principles of gravitational theories \cite{Cardoso:2018nvb,Destounis:2018qnb}.

The structure of this paper is outlined as follows. Sec. \ref{sec2} presents the field equation of the scalar perturbation in the EMS theory, reviews the violation condition of the SCC, and discusses the non-minimal couplings between the electromagnetic and scalar fields. In Sec. \ref{sec3}, we describe the numerical methods we used to compute the quasinormal modes (QNM) frequency, and we present the corresponding results. Finally, we summarize and discuss our findings in Sec. \ref{sec5}, which includes the identification of a critical coupling constant above which the SCC is reinstated for all black holes in RNdS spacetime.

\begin{table*}[ht]
    \centering
    \renewcommand\arraystretch{1.7}
    \begin{tabular}{|c|c|c|c|c|c|}
    \hline
    \makebox[0.1\textwidth][c]{ }& \makebox[0.12\textwidth][c]{$l=0$} & \makebox[0.12\textwidth][c]{$l=1$} & \makebox[0.12\textwidth][c]{$l=2$} & \makebox[0.12\textwidth][c]{$l=10$} & \makebox[0.12\textwidth][c]{$l=20$}\\
    \hline
    Pseudospectral method & $0.07967913$ & $0.06393574$ & $0.06182984$ & $0.06075811$ & $0.06071149$\\
    \hline
    Direct integration method & $0.07967913$ & $0.06393574$ & $0.06182984$ & $0.06075811$ & $0.06071149$\\
    \hline
    WKB  approximation&  &  & & $0.06073984$ & $0.06070675$\\
    \hline
    \end{tabular}
    \caption{The lowest-lying QNMs  $\beta=-{\text{Im}(\omega)}/{\kappa_-}$ with different $l$ calculated by different numerical methods for $\Lambda M^2 = 0.06, \alpha = 0.1$ and $Q/Q_\text{ext} = 0.9$.}\label{tb1}
\centering
    \renewcommand\arraystretch{1.7}
    \begin{tabular}{|c|c|c|c|c|c|}
    \hline
    \makebox[0.1\textwidth][c]{ }& \makebox[0.12\textwidth][c]{$l=0$} & \makebox[0.12\textwidth][c]{$l=1$} & \makebox[0.12\textwidth][c]{$l=2$} & \makebox[0.12\textwidth][c]{$l=10$} & \makebox[0.12\textwidth][c]{$l=20$}\\
    \hline
    Pseudospectral method & $0.57893700$ & $0.45043291$ & $0.43826722$ & $0.43203320$ & $0.43124841$\\
    \hline
    Direct integration method & $0.57893700$ & $0.45043291$ & $0.43826723$ & $0.43203451$ & $0.43176229$\\
    \hline
    WKB  approximation&  &  & & $0.43209053$ & $0.43177729$\\
    \hline
    \end{tabular}
    \caption{The lowest-lying QNMs  $\beta=-{\text{Im}(\omega)}/{\kappa_-}$ with different $l$ calculated by different numerical methods for $\Lambda M^2 = 0.06, \alpha = 0.1$ and $Q/Q_\text{ext} = 0.99$.}\label{tb2}
\end{table*}
\section{Strong cosmic censorship and quasinormal mode in Einstein-Maxwell-Scalar theory}\label{sec2}
In this paper, we investigate the Einstein-Maxwell-Scalar (EMS) theory, which involves a non-minimal coupling between a massless scalar field $\Phi$ and a Maxwell invariant term. The full action of this theory, given by

\begin{equation}\begin{aligned}\label{action}
S = \frac{1}{16\pi}\int &d^4 x\sqrt{-g}\left[R-2\Lambda\right.\\
&\left.-2\nabla_a \Phi\nabla^a \Phi-f(\Phi)F_{ab}F^{ab}\right],
\end{aligned}\end{equation}
includes the cosmological constant $\Lambda$, the Ricci scalar $R$, and the electromagnetic tensor $F_{ab}$. The coupling function $f(\Phi)$ determines the strength of the non-minimal coupling of $\Phi$ to the Maxwell term. To study the scalar field perturbation under first-order approximation, we require $f'(0)=0$, which can be achieved using a quadratic coupling function $1+\alpha \Phi^2$, an exponential coupling function $e^{\alpha \Phi^2}$, or other forms that satisfy
\begin{equation}
f(0)=1\,\quad \text{and}\quad f''(0)=2\alpha\,,
\end{equation}
where $\alpha$ is the coupling constant. The equations of motion, given by
\be\begin{aligned}\label{eomall}
&G_{ab}+\Lambda g_{ab}=T_{ab}^{\text{sc}}+T_{ab}^{\text{EM}}\,,\\
&\nabla^2\Phi-\frac{f'(\Phi)}{4} F_{ab}F^{ab}=0\,,\\
&\nabla^a [f(\Phi)F_{ab}]=0\,,
\end{aligned}\ee
are obtained by varying the action and consist of the Einstein equation, the scalar field equation, and the electromagnetic field equation. The energy-momentum tensor of the scalar field and the electromagnetic field are denoted by $T_{ab}^{\text{sc}}$ and $T_{ab}^{\text{EM}}$, respectively, and are defined by
\be\begin{aligned}\label{Tsc}
T_{ab}^{\text{sc}}&=2\nabla_a\Phi\nabla_b\Phi-g_{ab}\nabla_c\Phi\nabla^c\Phi\,,\\
T_{ab}^{\text{EM}}&=2f(\Phi)F_a{}^{c}F_{bc}-\frac{f(\Phi)}{2}F_{cd}F^{cd}g_{ab}\,.
\end{aligned}\ee
The EMS theory allows for the formation of scalarized black hole solutions through the non-minimal coupling between the scalar field and Maxwell invariant. By choosing an appropriate coupling function in the EMS theory, the electromagnetic vacuum solution, namely the RNdS solution, can become unstable under perturbations, resulting in the formation of scalarized black hole solutions. However, recent literatures \cite{Cai:2020wrp, An:2021plu} have shown that nonlinear electrodynamics black holes with scalar hairs, which lack a Cauchy horizon, effectively satisfy SCC. Therefore, the SCC problem only needs to be tested in the perturbatively stable black hole solutions, namely the RNdS solutions, which is given by
\be\begin{aligned}\label{metric}
ds^2&=-f(r)dt^2+\frac{1}{f(r)}dr^2+r^2 d\Omega^2\,,\\
A_a&=-\frac{Q}{r} (dt)_a\,,
\end{aligned}\ee
with the blackening factor
\be\begin{aligned}\label{blackening1}
f(r)=1-\frac{2M}{r}+\frac{Q^2}{r^2}-\frac{\Lambda r^2}{3}\,,
\end{aligned}\ee
describes a special solution of the EMS theory with vanishing scalar field.

Assuming $r_c$, $r_+$ and $r_-$ are the cosmological horizon, event horizon and Cauchy horizon respectively, we can rewrite the blackening factor as:
\begin{equation}
f(r)=\frac{\Lambda}{3r^2}(r_c-r)(r-r_+)(r-r_-)(r-r_o),
\label{blackening2}
\end{equation}
where $r_o$ is the minimum root of $f(r)=0$ and can be found to be $r_o=-r_c-r_+-r_-$. The surface gravity of each horizon can be defined as:
\begin{equation}
\kappa_i=\frac{1}{2}|f'(r_i)|\quad\text{with}\quad i={c,+,-,o}.
\end{equation}

We can expand the coupled scalar field $\Phi(t,r,\theta,\phi)$ on the RNdS background spacetime as a perturbation \cite{Kokkotas:1999bd,Berti:2009kk,Konoplya:2011qq}. Using the symmetries of the spacetime, we can express the scalar field in terms of spherical harmonics as
\begin{equation}\label{sepvar}
\Phi(t,r,\theta,\phi)=\sum_{lm}e^{-i \omega t}Y_{lm}(\theta,\phi)\frac{\psi(r)}{r}\,,
\end{equation}
where $Y_{lm}(\theta,\phi)$ is the spherical harmonics. By substituting Eqs. \eqref{metric} and \eqref{sepvar} into the equation of motion \eqref{eomall} satisfied by the scalar field, we obtain a one-dimensional Schr$\ddot{\text{o}}$dinger-like equation
\begin{equation}\label{schr}
\frac{d^2 \psi(r)}{dr^2_{\ast}}+[\omega^2-V(r)] \psi(r)=0\,,
\end{equation}
where the effective potential is given by
\begin{equation}\label{potential}
V(r)=\frac{f(r)}{r^2}\left[l(l+1)+r f'(r)-\frac{\alpha Q^2}{r^2}\right]\,,
\end{equation}
with the tortoise coordinate
\begin{equation}
dr_{\ast}=\frac{dr}{f(r)}\,.
\end{equation}
In the physical region between the event horizon $r_+$ and cosmological horizon $r_c$, the tortoise coordinate $r_{\ast}$ can be expressed as a sum of logarithmic terms involving the surface gravity of each horizon $i$, i.e.,
\begin{equation}\label{rs}\begin{aligned}
r_{\ast}=&-\frac{1}{2\kappa_c} \ln\left(1-\frac{r}{r_c}\right)+\frac{1}{2\kappa_+}\ln\left(\frac{r}{r_+}-1\right)\\
&-\frac{1}{2\kappa_-}\ln\left(\frac{r}{r_-}-1\right)+\frac{1}{2\kappa_o}\ln\left(1-\frac{r}{r_o}\right)\,.
\end{aligned}\end{equation}
Notably, the effective potential \eqref{potential} vanishes on every horizon, implying that the asymptotic solution of the Schr$\ddot{\text{o}}$dinger-like equation \eqref{schr} near each horizon $i$ takes the form:
\begin{equation}\label{asymptotic}
\psi\sim e^{\pm i\omega r_{\ast}},,\quad r\rightarrow r_i\,,
\end{equation}
where $e^{i \omega r_{\ast}}$ represents the outgoing wave and the other one represents the ingoing wave. Physical considerations \cite{Konoplya:2011qq} dictate that there is only an ingoing wave near the event horizon $r_+$ and only an outgoing wave near the cosmological horizon $r_c$, which can be expressed as the following boundary conditions:
\be\begin{aligned}\label{boundarycon}
&\psi\sim e^{-i\omega r_{\ast}}\,,\quad r\rightarrow r_+\,,\\
&\psi\sim e^{i\omega r_{\ast}}\,,\quad\,\,\,\, r\rightarrow r_c\,.
\end{aligned}\ee

The QNMs are the solutions of Eq. \eqref{schr} that satisfy the boundary condition \eqref{boundarycon}, and their frequencies $\omega$ are discrete. The validity of the SCC for an asymptotically de-Sitter black hole is determined by whether its weak solution can be extended beyond the Cauchy horizon. It was shown in Refs. \cite{Dias:2018ynt,Cardoso:2017soq} that this is possible if and only if the solution is locally square integrable derivative, meaning that it satisfies \ba\b\equiv -\frac{\text{Im} (\w)}{\k_-} \geq 1/2\,.
\ea
As long as there is a QNM having $\beta$ less than $1/2$, the Cauchy horizon is unable to be extended, which ensures the validity of SCC. Hence, to examine whether SCC is valid, we only need to investigate the lowest-lying QNM.

\begin{figure*}
    \centering
\includegraphics[width=0.32\textwidth]{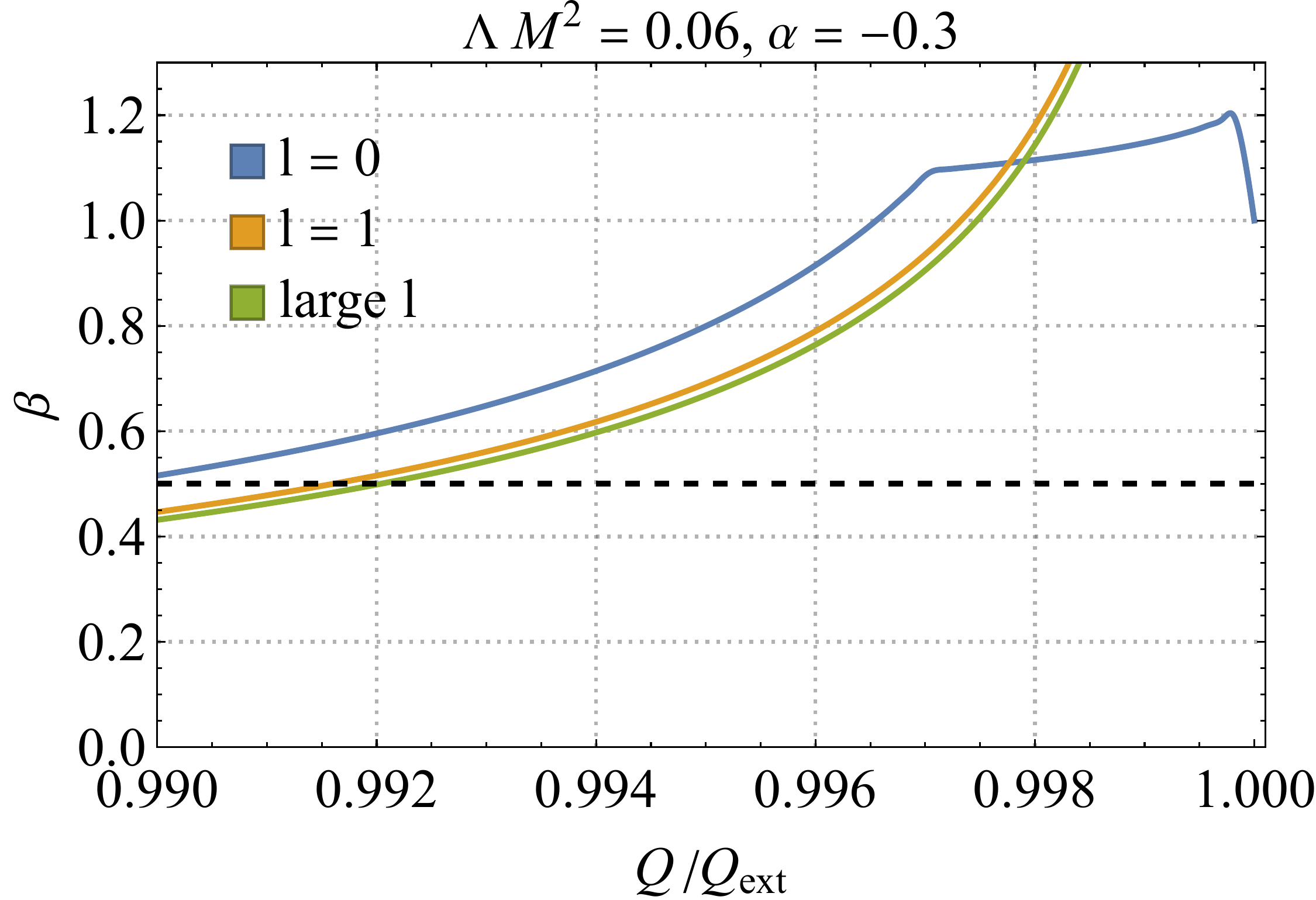}
 \includegraphics[width=0.32\textwidth]{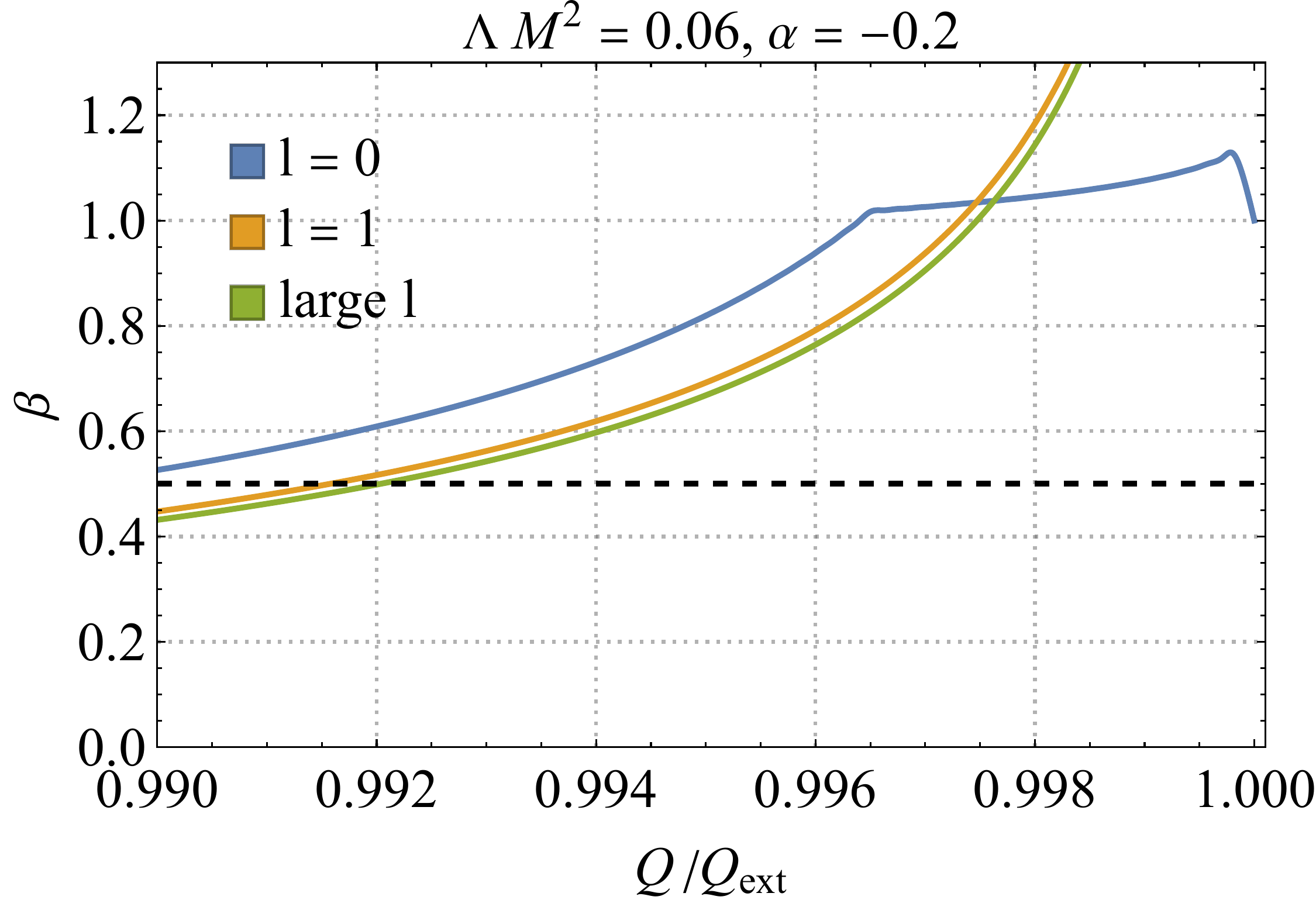}
 \includegraphics[width=0.32\textwidth]{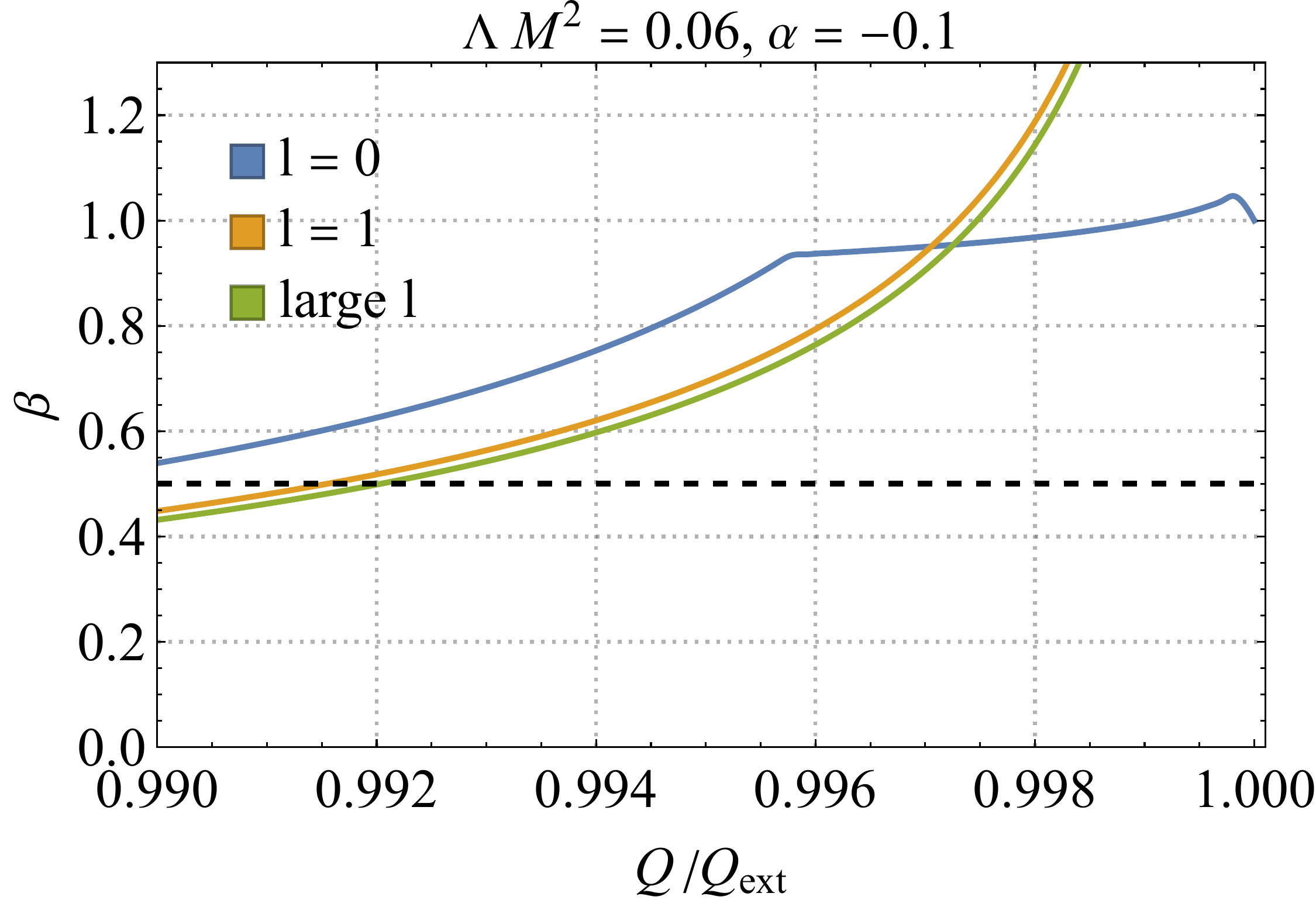}
 \includegraphics[width=0.32\textwidth]{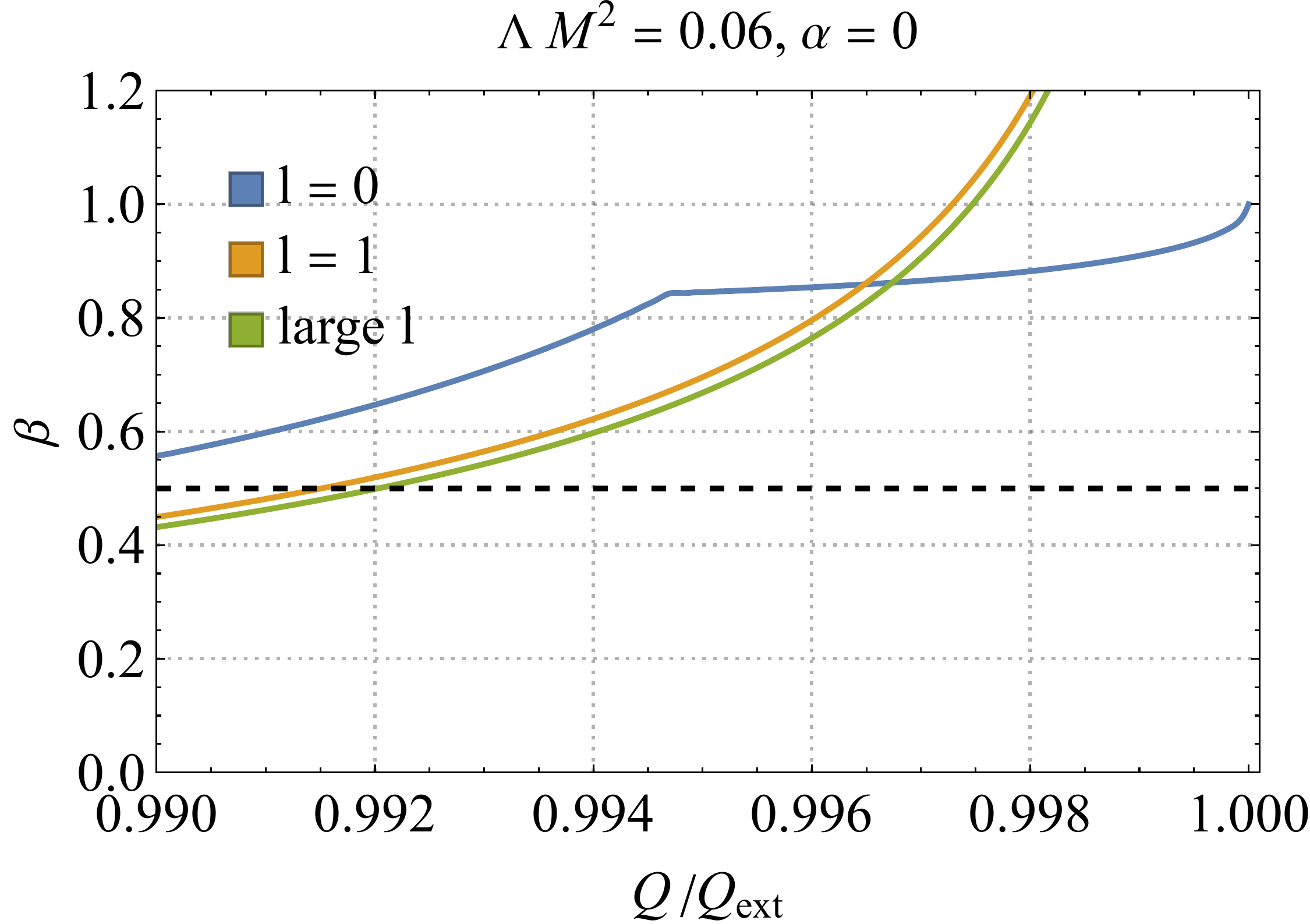}
 \includegraphics[width=0.32\textwidth]{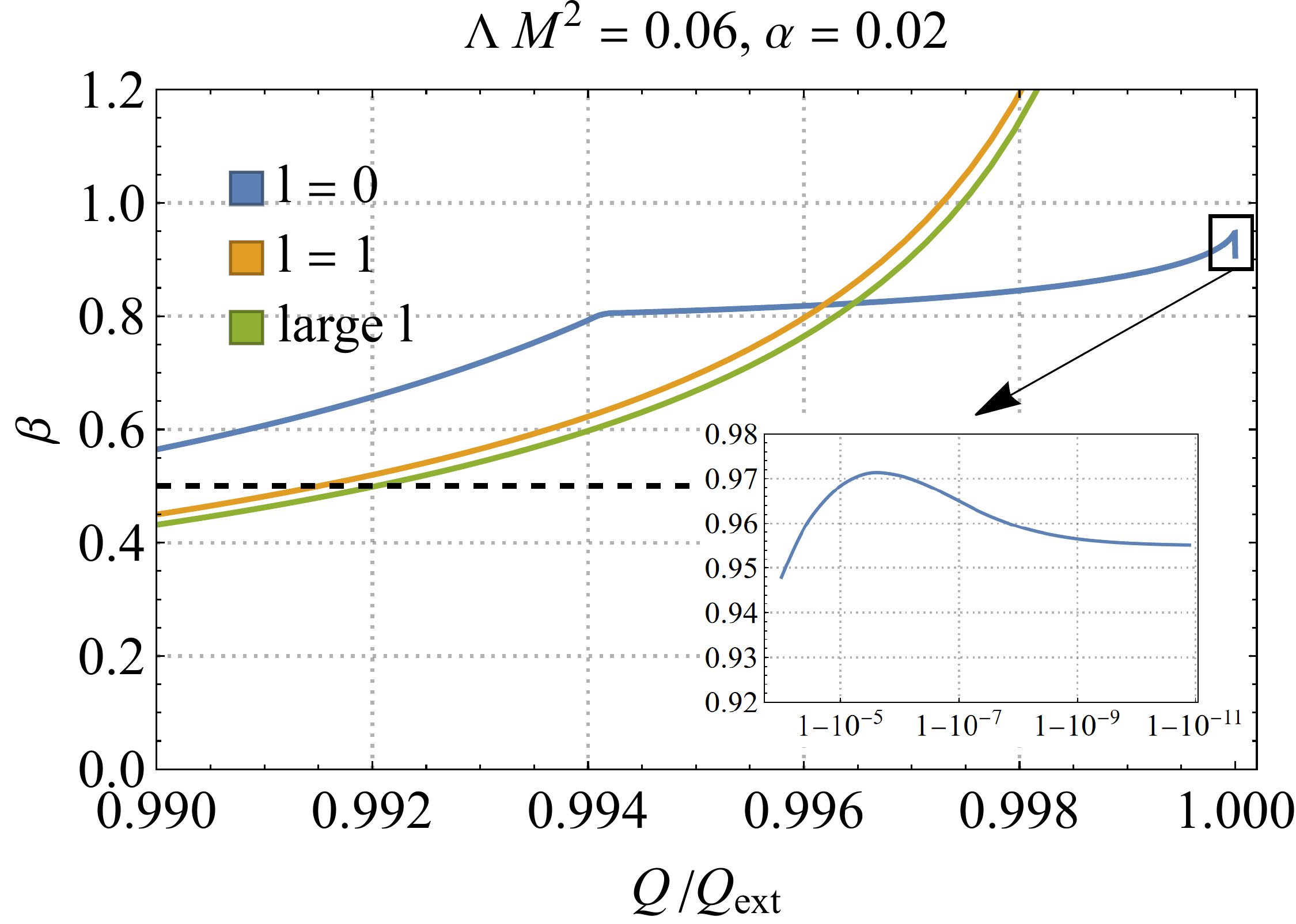}
 \includegraphics[width=0.32\textwidth]{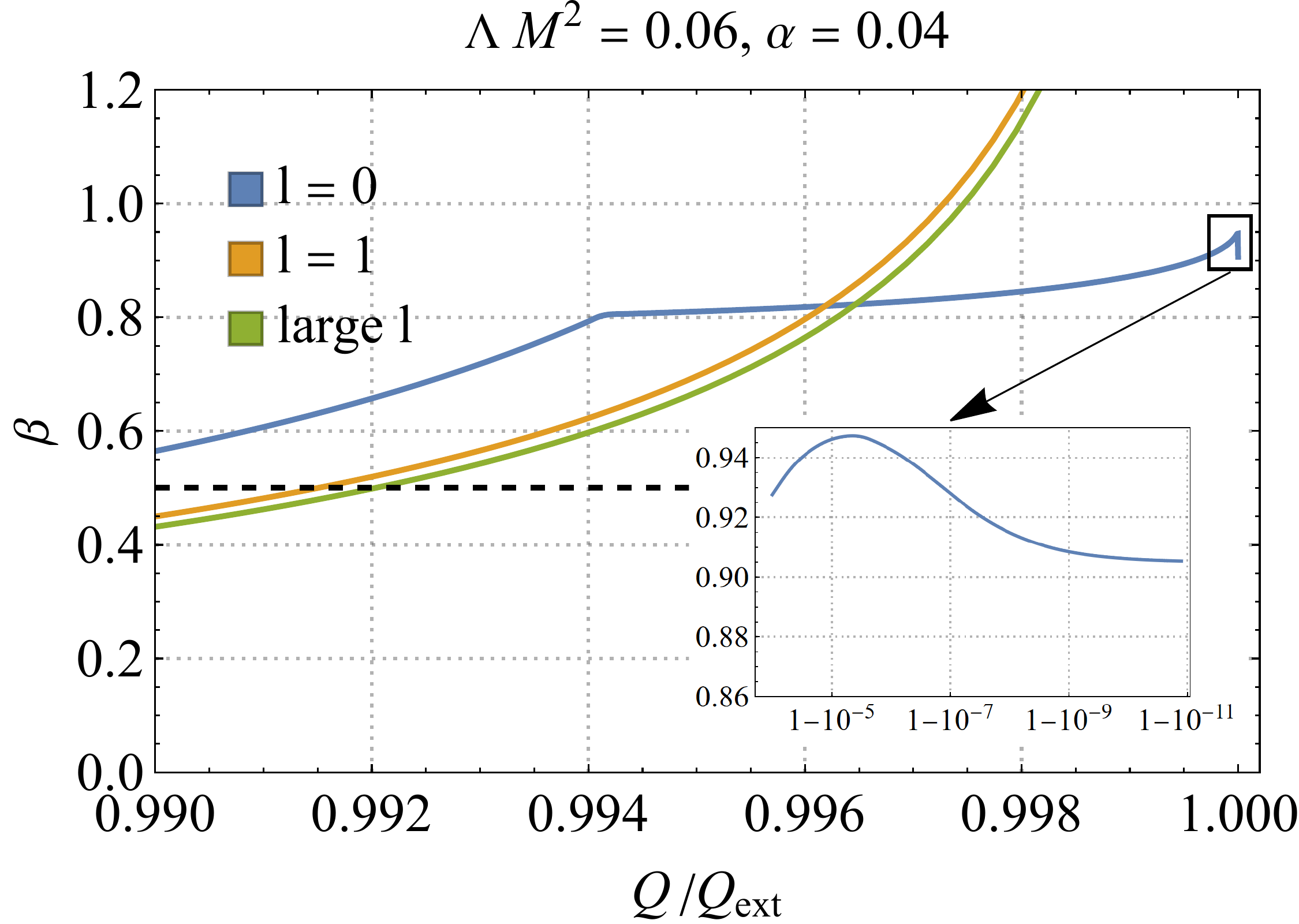}
  \includegraphics[width=0.32\textwidth]{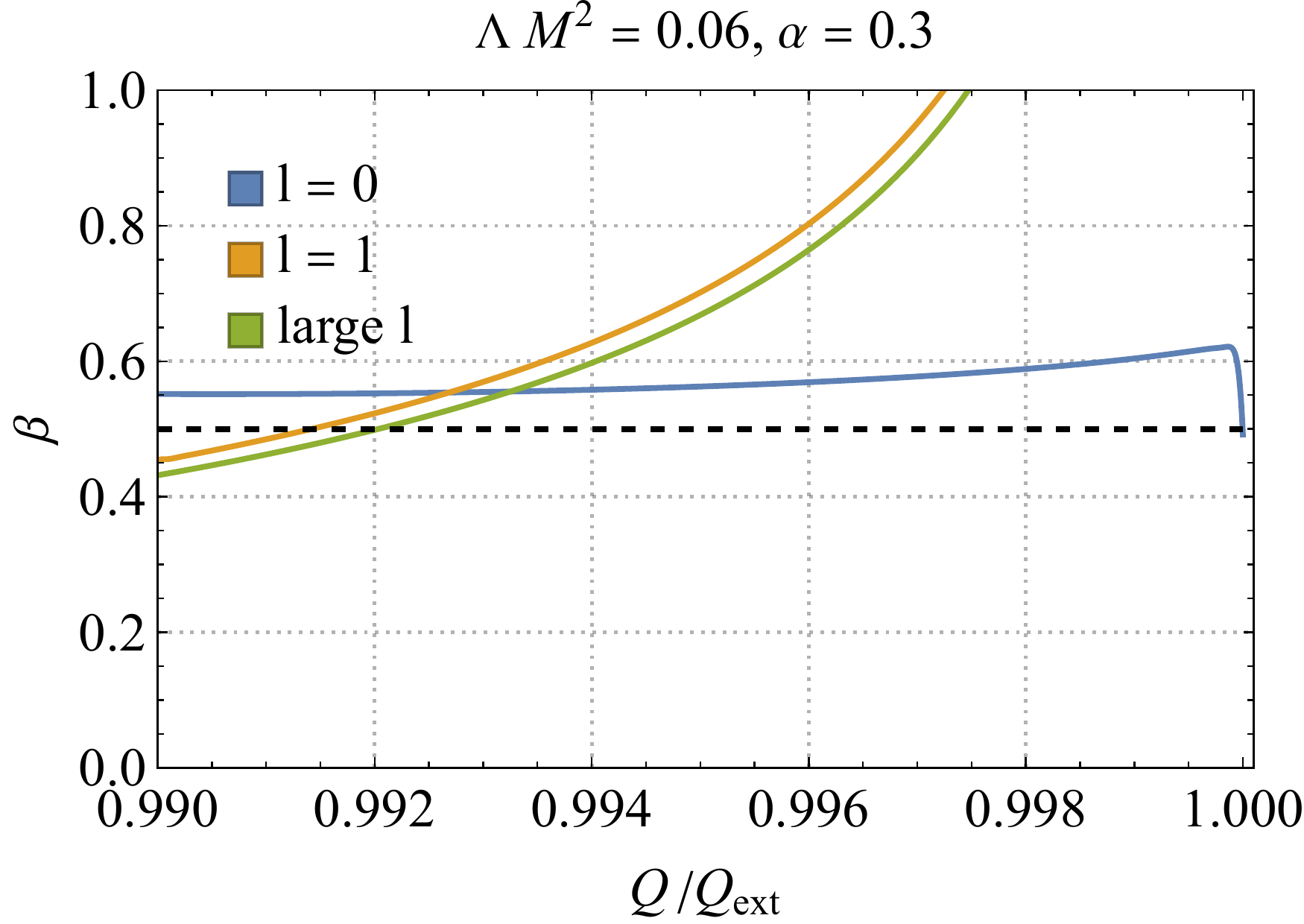}
   \includegraphics[width=0.32\textwidth]{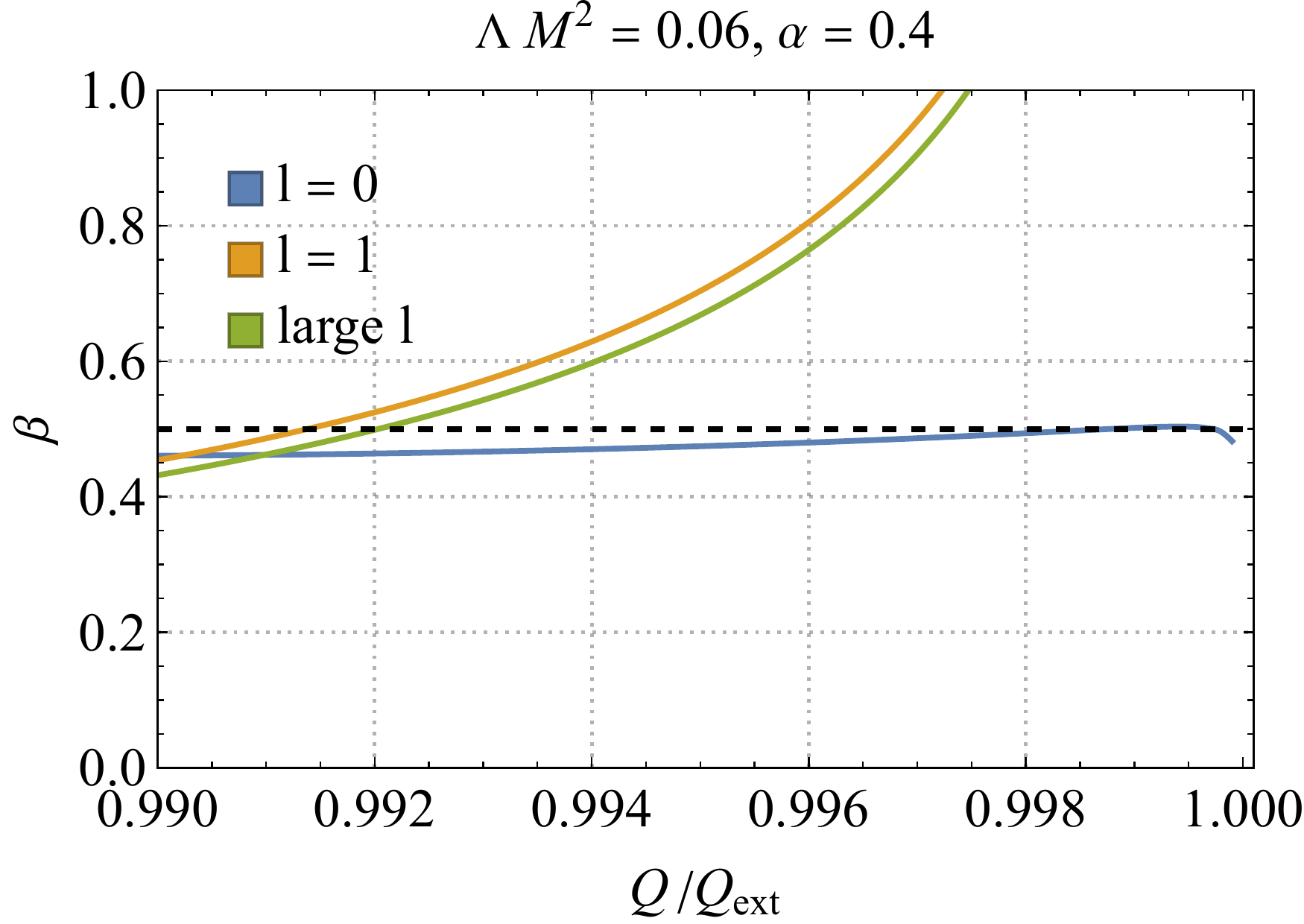}
    \includegraphics[width=0.32\textwidth]{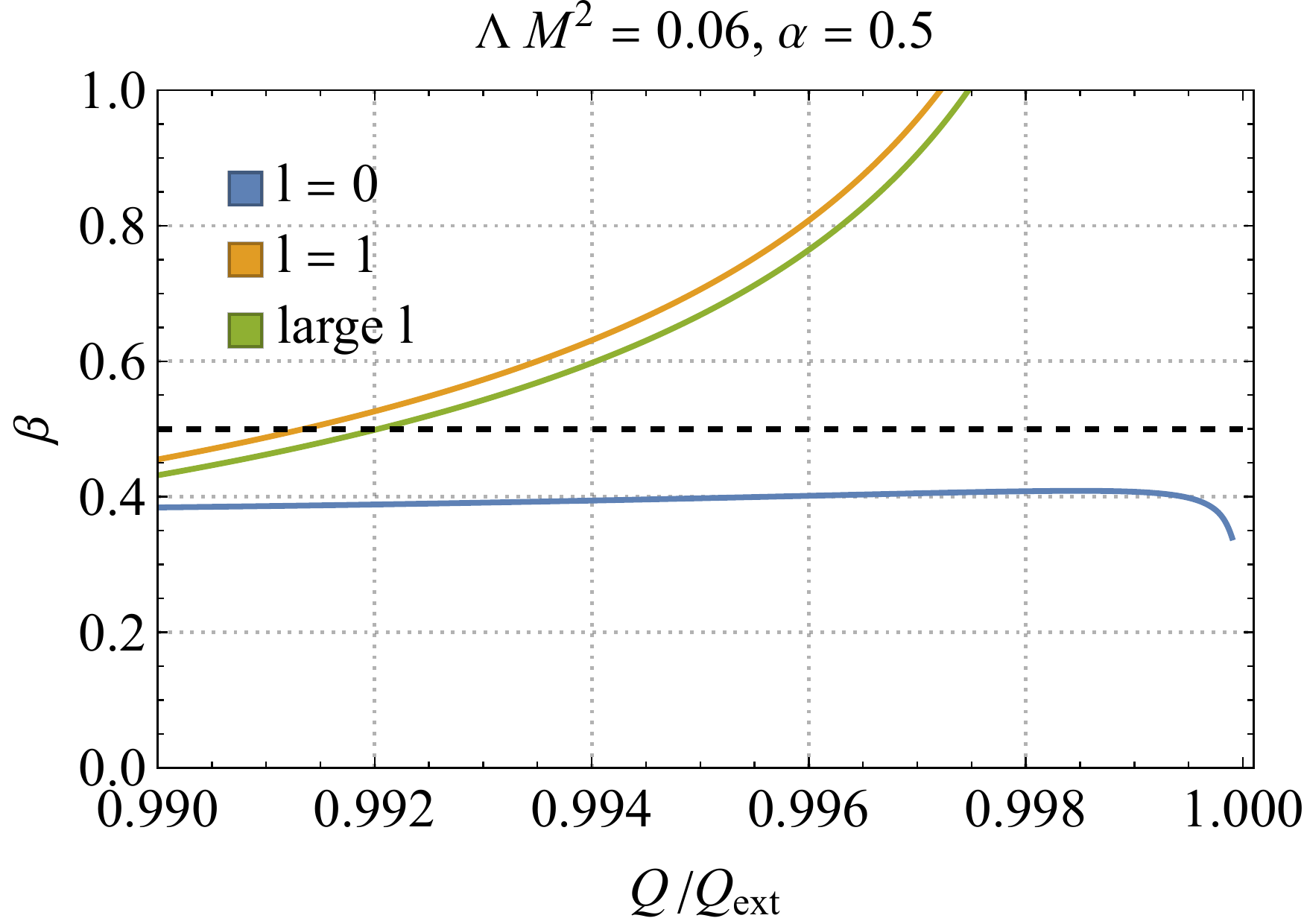}
    \caption{When $\L M^2 = 0.06$, the lowest-lying QNMs with the frequency $\beta=-{\text{Im}(\omega)}/{\kappa_-}$ are depicted for various coupling constants $\alpha$ as a function of the black hole charge ratio $Q/Q_\text{ext}$, for a given $l$. The SCC critical value of the charge ratio is indicated by the horizontal dashed line at $\b=1/2$.}
    \label{fig1}
\end{figure*}
\section{Numerical methods and results}\label{sec3}

In this section, we will present two novel numerical techniques to accurately determine the QNM frequencies.

While several numerical approaches have been developed for computing QNM frequencies with high precision \cite{Konoplya:2011qq}, in this study, we will introduce a pseudospectral method \cite{Jansen:2017oag,Miguel:2020uln} for calculating the QNM frequencies. To ensure the accuracy of our results, we will also validate them with the direct integration method \cite{Chandrasekhar:1975zza,Molina:2010fb}. Additionally, we will adopt the WKB approximation \cite{Konoplya:2003ii} to calculate the photo-sphere modes, which correspond to the QNMs in the large-$l$ limit.

We notice that the scalar field $\y(r)$ oscillates significantly near the two horizons, namely the event horizon and the cosmological horizon, as given in equations \eqref{boundarycon}. To apply the pseudospectral method to the field equation \eqref{schr}, we introduce a new variable $y(x)$ defined by
\ba\begin{aligned}\label{respsi}
\psi(r)=\left(x+1\right)^{\frac{i \omega}{2\kappa_+}}\left(x-1\right)^{-\frac{i\omega}{2\kappa_c}}y(x)\,.
\end{aligned}\ea
to transform the field equation into a regular form in the interval $[-1,1]$. In particular, the variable $x$ is defined as a function of $r$ using the relationship between $r$ and $x$ given in the equation below,
\begin{equation}
r = \frac{r_c-r_+}{2}x+\frac{r_c+r_+}{2},.
\end{equation}
After this transformation, the field equation \eqref{schr} can be written as
\ba\begin{aligned}\label{speeqy}
a_0(\omega,x)y(x)+a_1(\omega,x)y'(x)+a_2(\omega,x)y''(x)=0\,,
\end{aligned}\ea
which is a second-order differential equation with variable coefficients.

To apply the pseudospectral method, we first expand the field equation and variable $y(x)$ by the cardinal function $C_i(x)$, which satisfies $C_i(x_j)=\delta_{ij}$, where $x_j$ denotes the $j$-th Gauss-Lobatto point. Substituting this expansion into equation \eqref{speeqy} and multiplying each term by $C_k(x)$, we obtain a set of algebraic equations. By imposing the boundary conditions given in equation \eqref{boundarycon}, we can obtain a matrix equation of the form $(M_0+\omega M_1) Y=0$, where $M_0$ and $M_1$ are matrices and $Y$ is a vector containing the coefficients of the expansion. Then, the QNM frequency can be obtained by solving the eigenvalue problem of the matrix $(-M_1^{-1}M_0)$. This method allows us to calculate the QNM frequencies for scalar perturbations on the RNdS black hole background with high precision.

\begin{figure*}
    \centering
\includegraphics[width=0.48\textwidth]{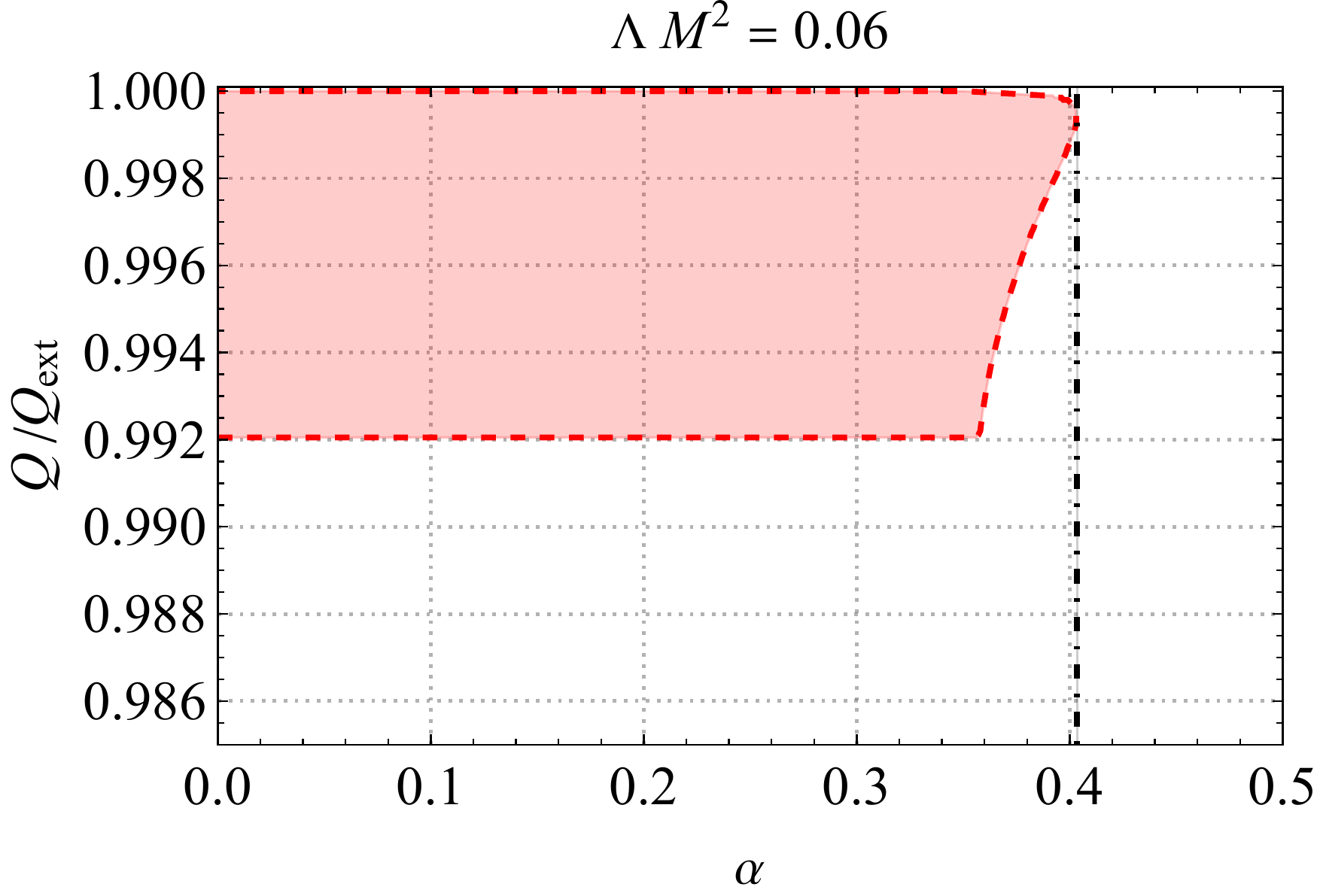}
 \includegraphics[width=0.48\textwidth]{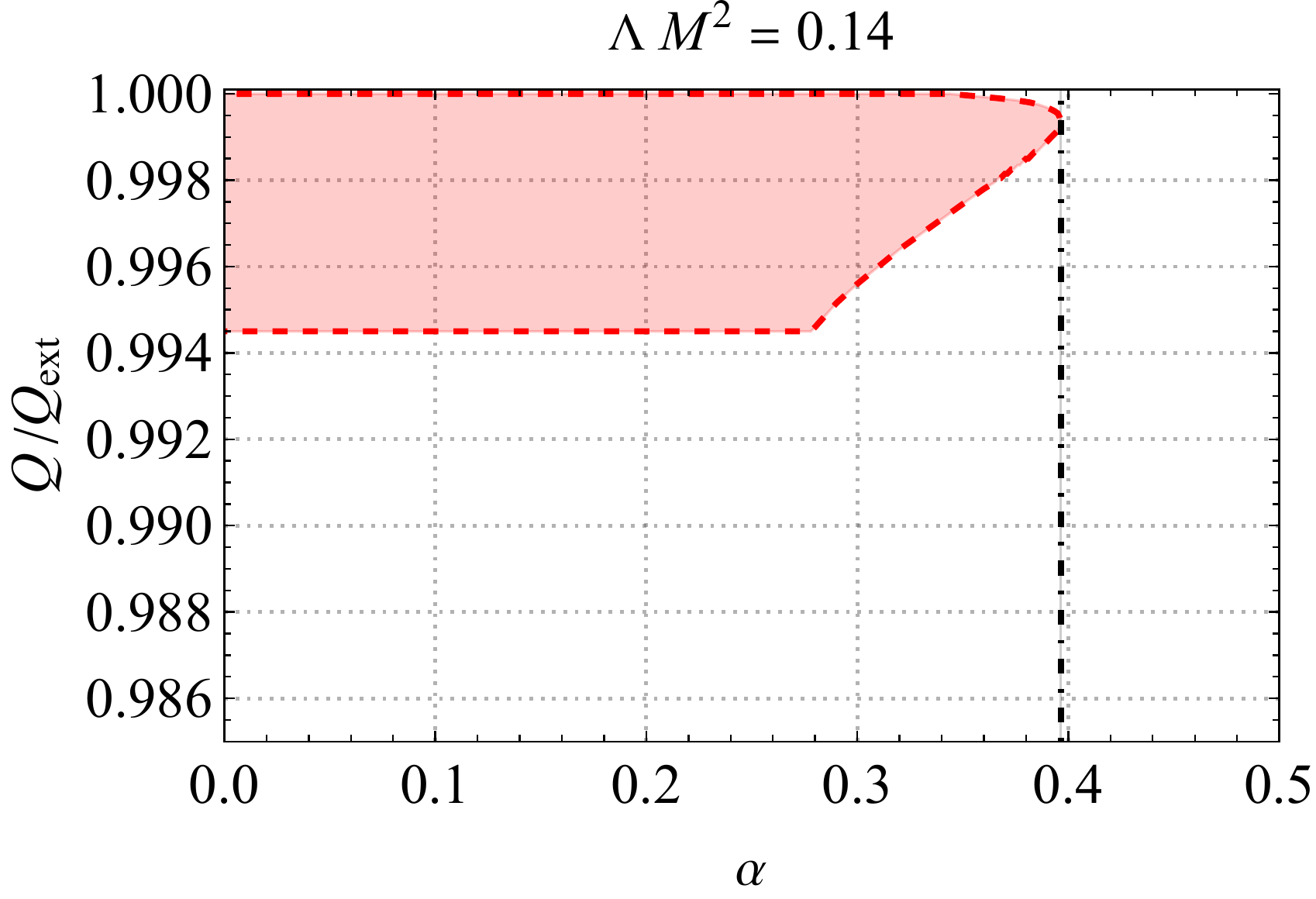}
    \caption{The red shaded region in the $\alpha-Q/Q_\text{ext}$ diagram represents the region where the SCC is violated for $\L M^2=0.06$ and $\L M^2=0.14$, respectively. The critical value of $\alpha = \alpha_\text{crit}$, represented by the black dot-dashed vertical line, indicates the point where all black holes satisfy the SCC for EMS theory with a coupling constant $\alpha > \alpha_\text{crit}$.}
    \label{fig2}
\end{figure*}

Tables \ref{tb1} and \ref{tb2} display the QNM frequencies obtained from both the pseudospectral method and the direct integration method. For the direct integration method, we used the series expansion of $y(x)$ near $x=\pm1$ as the boundary condition and solved equation \eqref{speeqy} in the intervals $(-1,0]$ and $[0,1)$. This was done for a given $\omega$ using \emph{Mathematica}, and the acceptable frequency $\omega$ was determined by ensuring the two solutions were smooth at $x=0$.

We have utilized both the pseudospectral method and direct integration method to calculate the lowest-lying QNMs $\b = -\text{Im} (\w)/\kappa_-$ for different $l$ and various black hole parameters. The results are presented in Tables \ref{tb1} and \ref{tb2}, which demonstrate the reliability of our numerical computations. Here, $Q_\text{ext}$ represents the electric charge of the extremal black hole solution, which can be derived from the condition $r_+ = r_-$. Furthermore, we have employed the WKB approximation to evaluate the large-$l$ lowest-lying modes, and our results are consistent with other methods. Therefore, the lowest-lying modes for large $l$ are determined by the WKB approximation in our study. It has been previously established that the SCC is hardly violated for the RNdS black holes that are not near extremal. Our results for the EMS theory are consistent with previous research. Hence, we only present the QNMs results in the nearly extremal region in this paper.

Fig. \ref{fig1} displays the lowest-lying QNMs with frequency $\beta=-{\text{Im}(\omega)}/{\kappa_-}$ for different coupling constants $\alpha$ as a function of the black hole charge ratio $Q/Q_\text{ext}$, at a given $l$. The graphs reveal that, for most coupling constants, there is a range of charge parameters near the black hole's extremal limit where all QNMs violate the SCC by having $\beta > 1/2$. We can determine the violation region of the SCC by identifying the intersection point of the $\beta=1/2$ curve and the curve of all lowest-lying modes for all possible values of $l$. However, at specific values of the coupling constant, such as $\alpha = 0.5$, all black holes satisfy the SCC, meaning that all lowest-lying QNMs for all $l$ are less than $1/2$. Moreover, the lowest-lying QNMs with $l=0$ (nearly extremal modes) contribute the most to the preservation of the SCC. The plots reveal that the lowest-lying modes correspond to either $l=0$ or large $l$. The top three plots represent cases where the coupling constant is negative: $\alpha = -0.1$, $\alpha = -0.2$, and $\alpha = -0.3$. As the coupling constant becomes more negative, the lowest-lying QNMs with $l=0$ increase, indicating that restoring the SCC in these cases requires increasing the coupling constant in the positive direction. This is expected because a larger coupling constant $\alpha$ leads to a smaller effective potential, which can even become negative and destabilize the spacetime under scalar perturbations.

Finally, we investigate the impact of the non-minimal coupling constant $\alpha$ on the validity of the SCC by plotting the violation region of the SCC in the $\alpha-Q/Q_\text{ext}$ diagram (see Figs. \ref{fig2}) for two different values of $\Lambda M^2$: 0.06 and 0.14. We exclude the region where $\alpha$ is negative since it extends to minus infinity and can be inferred from the positive $\alpha$ region. Our analysis demonstrates that for small positive coupling constants $\alpha$, the SCC is violated as the black hole approaches the extremal limit. However, as $\alpha$ increases, the SCC can be partially or fully restored, depending on the value of $\alpha$. We identify a critical value $\alpha_\text{crit}$ of the coupling constant, above which the SCC is reinstated for all black holes in RNdS spacetime, regardless of their charge-to-mass ratio. In other words, the validity of the SCC no longer depends on the charge-to-mass ratio of the black hole in these situations. Thus, we can say that this theory satisfies the SCC. Our results provide valuable insights into the restoration of the SCC in RNdS spacetime and suggest that the non-minimal couplings between the electromagnetic and scalar fields may have an important role in the process.

\section{conclusion and discussion}\label{sec5}

In this paper, we have investigated the validity of the SCC in the context of RNdS black holes within the framework of the EMS theory. We have examined the impact of scalar field perturbations on the SCC in RNdS spacetime with non-minimal couplings between the electromagnetic and scalar fields.

Our numerical analysis of the QNM frequencies of the non-minimally coupled scalar field reveals that the SCC can be reinstated within a specific range of the coupling constant $\alpha$. By plotting the violation region of the SCC in the $\alpha-Q/Q_\text{ext}$ diagram for two different values of $\Lambda M^2$, we have identified a critical coupling constant $\alpha_\text{crit}$, which enables the EMS theory with $\alpha > \alpha_\text{crit}$ to satisfy the SCC. Our results provide new insights into the restoration of the SCC in RNdS spacetime and suggest that non-minimal couplings between the electromagnetic and scalar fields may play an important role in the restoration of the SCC.

Our findings have important implications for other modified gravitational theories, and it would be valuable to explore whether similar restoration of the SCC can be achieved in those theories. Additionally, our analysis has focused on the linear level of the EMS theory, and it would be worthwhile to extend our investigation to the non-linear regime. Finally, it would be interesting to explore the physical implications of the restoration of the SCC in the context of astrophysical black holes and the evolution of the universe.

\section*{Acknowledgement}

Jie Jiang is supported by the National Natural Science Foundation of China with Grant No. 12205014, the Guangdong Basic and Applied Research Foundation with Grant No. 217200003 and the Talents Introduction Foundation of Beijing Normal University with Grant No. 310432102. Jia Tan is supported by the starting funding of Suzhou University of Science and Technology with Grant no. 332114702, Jiangsu Key Disciplines of the Fourteenth Five-Year Plan with Grant No. 2021135, and Natural Science Foundation of JiangSu Province (BK20220633).

\end{document}